\documentclass{article}
\usepackage{spconf}
\usepackage{amsmath}
\usepackage{graphicx}
\usepackage{array}
\usepackage{url}
\usepackage{balance}
\usepackage{bbm}
\usepackage{cite}
\usepackage{amssymb}
\usepackage{xcolor}
\usepackage{multirow,enumitem,adjustbox}
\usepackage{booktabs,threeparttable,makecell}
 \usepackage[colorlinks=true,citecolor=green]{hyperref}
\usepackage{cleveref}

\title{Flow-TSVAD: Target-Speaker Voice Activity Detection via Latent \\ Flow Matching}

\name{Zhengyang Chen$^{1}$, Bing Han$^{1}$, Shuai Wang$^{2,3}$, Yidi Jiang$^{4}$, Yanmin Qian$^{1}$}

\address{
$^1$AudioCC Lab, CS Dept, Shanghai Jiao Tong University, Shanghai, China\\
$^{2}$Shenzhen Research Institute of Big Data, Shenzhen, China \\
$^{3}$School of Data Science, The Chinese University of Hong Kong, Shenzhen, China \\
$^{4}$National University of Singapore, Singapore
  }
 
\begin{document}
\ninept
\topmargin=0mm
\maketitle

\begin{abstract} 
Speaker diarization is typically considered a discriminative task, using discriminative approaches to produce fixed diarization results. In this paper, we explore the use of neural network-based generative methods for speaker diarization for the first time. We implement a Flow-Matching (FM) based generative algorithm within the sequence-to-sequence target speaker voice activity detection (Seq2Seq-TSVAD) diarization system. Our experiments reveal that applying the generative method directly to the original binary label sequence space of the TS-VAD output is ineffective. To address this issue, we propose mapping the binary label sequence into a dense latent space before applying the generative algorithm and our proposed Flow-TSVAD method outperforms the Seq2Seq-TSVAD system. Additionally, we observe that the FM algorithm converges rapidly during the inference stage, requiring only two inference steps to achieve promising results. As a generative model, Flow-TSVAD allows for sampling different diarization results by running the model multiple times. Moreover, ensembling results from various sampling instances further enhances diarization performance.

\end{abstract}

\section{Introduction}
Speaker diarization system aims to answer the question ``who spoke when", which is crucial for multi-speaker speech processing scenarios. Typically employed as a pre-processing technique, speaker diarization's output can be used by automatic speech recognition (ASR) systems to generate speaker-attributed transcriptions~\cite{kanda2019simultaneous}. In the era of big data and large models, it also serves as an effective tool for filtering out single-speaker utterances from noisy internet-collected data~\cite{ma2024wenetspeech4tts}.

The speaker diarization system has undergone a long period of development, transitioning from clustering-based methods to neural network-based approaches. In clustering-based methods, Agglomerative Hierarchical Clustering (AHC)~\cite{tranter2006overview}, Spectral Clustering (SC)~\cite{wang2018speaker}, or Variational Bayesian clustering (VBC)~\cite{valente2010variational,landini2022bayesian} is used to cluster the segment-level speaker representation. However, clustering algorithms can assign only one label to each segment, making them incapable of handling speaker overlap. To address this limitation, neural network-based methods have emerged, primarily including two paradigms: end-to-end neural diarization (EEND)~\cite{fujita2019end_lstm,fujita2019end_sa,horiguchi2022encoder,chen23n_interspeech,chen2024attention} and target speaker voice activity detection (TS-VAD)~\cite{medennikov2020target,chen23n_interspeech,cheng2023target,chen2024attention,cheng2024multi}. EEND systems take audio features as input and directly output diarization results for each speaker. In contrast, TS-VAD systems use both audio features and speaker enrollment embeddings as input, outputting diarization results for each enrolled speaker. TS-VAD systems often exhibit superior performance and are frequently used in winning competition systems~\cite{medennikov2020stc,wang2022dku}, attracting increasing research interest.

Although VBC can be considered a statistic generative method, neural network-based speaker diarization is usually considered a discriminative task, meaning that we want the network to provide a definite prediction. Unlike generative tasks, generative task-based systems often model the distribution of the data rather than producing a fixed output. Recently, generative algorithms have been applied to some speech-related tasks that previously used discriminative methods, such as speech enhancement~\cite{richter2023speech,welker22_interspeech} and target speaker extraction~\cite{kamo2023target}. These tasks should have deterministic ground truth output, but generative methods are increasingly used to improve the subjective evaluation of generated speech. Apart from speech tasks, generative methods have also been applied to some typically discriminative computer vision tasks, such as object detection~\cite{chen2023diffusiondet} and instance segmentation~\cite{gu2024diffusioninst}, resulting in performance improvements.

In this paper, we explore the application of neural network-based generative methods to speaker diarization for the first time. We implement a flow-matching-based generative algorithm~\cite{lipman2023flow} on the competitive sequence-to-sequence TS-VAD (Seq2Seq-TSVAD) system~\cite{cheng2023target,cheng2024multi}. Initial experiments showed that applying the generative method to the original binary label sequence space was ineffective. To address this, we propose using a pre-trained AutoEncoder (AE) to transform the binary label sequence into a dense latent space before applying the generative algorithm and our proposed Flow-TSVAD outperforms the strong Seq2Seq-TSVAD system. Additionally, as a generative model, running the Flow-TSVAD system multiple times produces varied results. By ensembling these different predictions, we can further improve the diarization performance.


\begin{figure*}[ht!]
    \centering
    \includegraphics[width=0.99\textwidth]{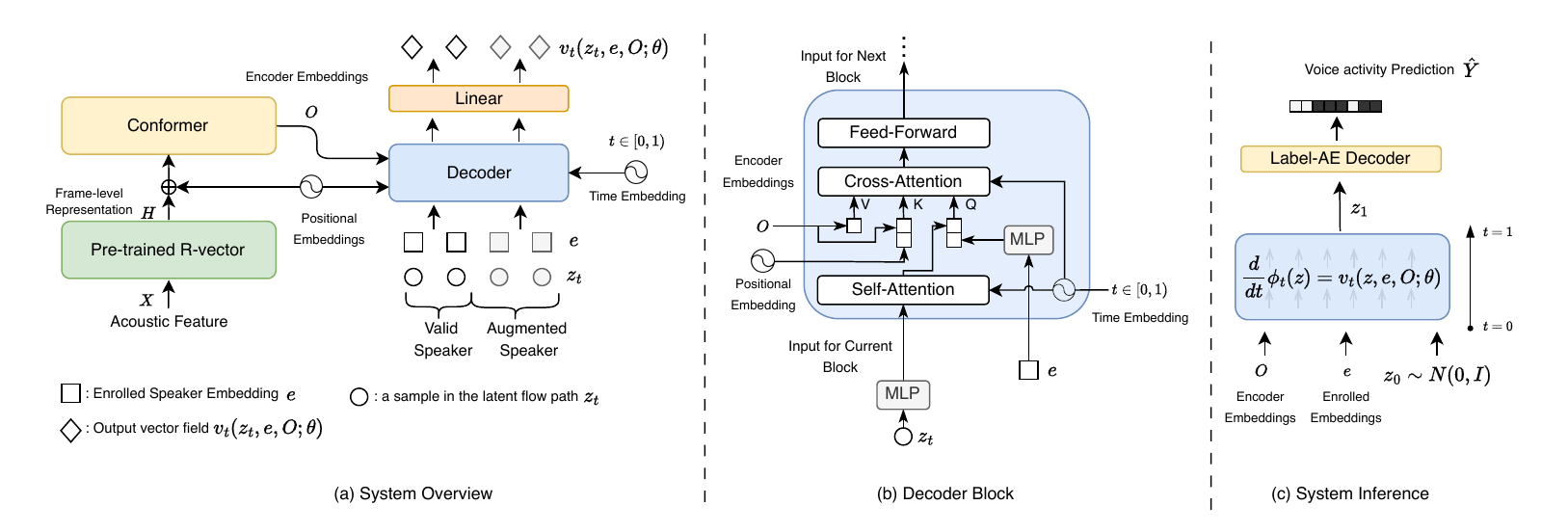}
    \caption{\textbf{System Overview of Flow-TSVAD system.}
    }
    \label{fig:system_overview}
\vspace{-10pt}
\end{figure*}

\section{Method}

\subsection{Flow Matching Algorithm}
\label{ssec:flow_matching_intro}

Deep learning techniques encounter significant challenges when modeling the distribution of data points $z_1 \in \mathbb{R}^d$ drawn from an unknown distribution $q(z_1)$. Generative models are typically tasked with learning the transformation from a simple prior distribution $p_0$ (e.g., Gaussian) to a target distribution $p_1 \approx q$. The flow matching algorithm \cite{lipman2023flow} addresses this by constructing a continuous flow $\phi_t: \mathbb{R}^d \rightarrow \mathbb{R}^d$ for $t \in [0, 1]$, which maps the prior to the target distribution via the regression of a vector field $u_t \in \mathbb{R}^d$. This transformation is governed by the following ordinary differential equation (ODE):
\begin{equation}
    \frac{d}{d t} \phi_t(z) = u_t\left(\phi_t(z)\right)
\end{equation}

To approximate the path of the flow, $u_t$ can be estimated using a neural network. However, due to the lack of a closed-form expression for $u_t$, direct approximation is challenging. Lipman et al.~\cite{lipman2023flow} propose substituting the original vector field $u_t$ with a conditional vector field $u_t(z|z_1)$, leading to the Conditional Flow Matching (CFM) objective:
\begin{equation}
\label{eq:CFM}
\mathcal{L}_{\mathrm{CFM}}(\theta) = \mathbb{E}_{t, q(z_1), p_t(z|z_1)}\left[\Vert v_t(z, \theta) - u_t(z|z_1)\Vert^2\right]
\end{equation}
Here, $p_t(z|z_1)$ denotes the conditional probability density function at time $t$, and $v_t(z, \theta)$ is the neural network used to approximate $u_t(z|z_1)$. Lipman et al.~\cite{lipman2023flow} demonstrate that approximating $u_t(z|z_1)$ is equivalent to approximating $u_t$.

The flow path can be characterized using the optimal transport (OT) path as described by \cite{lipman2023flow}, where $p_t(z|z_1) = \mathcal{N}(x| t z_1, (1 - (1 - \sigma_{\min}) t)^2 I)$ and $u_t(z|z_1) = (z_1 - (1 - \sigma_{\min}) z) / (1 - (1 - \sigma_{\min}) t)$. In this formulation, we set $\sigma_{\min}$ to zero value.

\subsection{VAD Label Auto-Encoder}
\label{ssec:label_ae}

The flow-matching (FM) algorithm operates on a continuous space. However, in neural network-based diarization systems, the speech activity for each speaker is always represented by a binary sequence $Y  = [y_1, y_2, ..., y_L] \in \{0, 1\}^L$, where $1$ represents the existence of the speaker,  while $0$ indicates the absence. We also believe that leveraging the FM algorithm in a discrete binary space cannot fully utilize its potential. Therefore, we propose a label auto-encoder (Label-AE) that compresses the labels in the binary space into a continuous and denser latent space. Generally speaking, for TS-VAD systems, we often divide long audio into fixed-length segments as input. This way, the labels for each segment are also of fixed length. Assuming we have a binary label sequence, $Y \in \{0, 1\}^L$, for one speaker in one segment, the Label-AE encoder maps the label sequence to $z_1 \in \mathbb{R}^{k}$, which resides in a lower and denser space. The Label-AE decoder reversely maps the latent vector $z_1$ into a $L$-length vector $\hat{Y}  = [\hat{y}_1, \hat{y}_2, ..., \hat{y}_L] \in [0, 1]^L$, where each value represents the speech probability at that time. To model the locality information of the speech activity label, we design the encoder and decoder by employing one-dimensional convolution. The detailed architecture of our proposed Label-AE is introduced in section \ref{ssec:model_configuration}. In the training process of the Label-AE, we directly optimize the binary cross-entropy loss between the $Y$ and $\hat{Y}$:
\begin{equation}
    \mathcal{L} = \sum_{i=1}^L\left[-y_i \log \hat{y}_i-\left(1-y_i\right) \log \left(1-\hat{y}_i\right)\right]
\end{equation}

\subsection{Generative TSVAD based on Sequence-to-Sequence Modeling}
\label{ssec:system_overview}

In this section, we propose incorporating the generative flow-matching (FM) algorithm into the TS-VAD system. To reduce the modeling complexity of the TS-VAD model, as illustrated in the left part of Figure \ref{fig:system_overview}, we construct our model using an encoder-decoder architecture, following the design of Seq2Seq-TSVAD~\cite{cheng2023target,cheng2024multi}. We denote the acoustic feature from the input audio as $X = [x_1, x_2, \ldots, x_{RL}] \in \mathbb{R}^{RL \times F}$. First, a pre-trained speaker encoder, r-vector \cite{zeinali2019but}, maps the input $X$ into the frame-level speaker representation $H = [h_1, h_2, \ldots, h_{L}] \in \mathbb{R}^{L \times E}$ with an $R$ times downsample rate. Then $H$ is added with sinusoidal position embeddings~\cite{vaswani2017attention} and fed into another Conformer~\cite{gulati2020conformer} module to obtain $O = [o_1, o_2, \ldots, o_{L}] \in \mathbb{R}^{L \times d}$.

As introduced in Section \ref{ssec:flow_matching_intro}, we approximate the vector field $u_t(z|z_1)$ to construct the flow path. To apply the FM algorithm in a dense latent space, we first map the VAD label $Y \in \{0, 1\}^L$ for each speaker to the latent vector $z_1 \in \mathbb{R}^{k}$. We then sample a timestamp $t$ from $[0, 1)$ and sample $z_t$ from $p_t(z|z_1)$ as introduced in Section \ref{ssec:flow_matching_intro}. The decoder takes $z_t$, the output from the encoder $O$, and the enrolled speaker embedding $e \in \mathbb{R}^c$ as input and outputs the approximated vector field $v_t(z_t, e, O; \theta) \in \mathbb{R}^{k}$ for each speaker. As illustrated in the middle part of Figure \ref{fig:system_overview}, the decoder block takes $z_t$ as the direct input and forwards it to the self-attention module. The enrollment embedding $e$ is used as conditioning information, concatenated with the query input for the cross-attention module. The positional embedding is also concatenated with the key value of the cross-attention module to incorporate the input's temporal information. Moreover, we replace the original layer-norm operation in the self-attention and cross-attention modules of the decoder with an Adaptive Instance Normalization (AdaIN)~\cite{huang2017arbitrary} to incorporate the time information used in the FM algorithm. Finally, a linear layer transforms the decoder's output to the predicted vector field $v_t(z_t, e, O; \theta)$. During the inference stage of diarization, illustrated by the right part of Figure \ref{fig:system_overview}, we obtain the VAD prediction in the latent space $z_1$ for each speaker by integrating the ODE function from $t=0$ to $t=1$:
\begin{equation}
    \frac{d}{d t} \phi_t(z) = v_t(z_t, e, O; \theta); \quad \phi_0(z) = z_0 \sim N(0, I)
\end{equation}
Finally, $z_1$ is mapped to the predicted voice activity $\hat{Y}$ using the Label-AE decoder.

\section{Experiment Setup}

\subsection{Dataset}
In our experiment, we use the CALLHOME~\cite{callhome} dataset for system evaluation, which contains utterances with number-of-speakers ranging from 2 to 7. The CALLHOME is split into two parts, part1 and part2, following the Kaldi recipe\footnote{\url{https://github.com/kaldi-asr/kaldi/tree/master/egs/callhome_diarization/v2}}. The part1 is used for model training, and the part2 for evaluation. To further expand the training data for the model,
we additionally simulate 1000 hours 3-speaker and 1000 hours 4-speaker data by leveraging the single speaker data from the NIST SRE dataset and Switchboard dataset introduced in \cite{landini22_interspeech} following the simulation recipe in~\cite{landini22_interspeech}. In the simulation process, the statistics from the CALLHOME part1 are used.

\begin{table}[ht!]
\footnotesize
\centering
\caption{\textbf{Model architecture for Label-AE.} For rows containing Conv1D, the numbers in brackets indicate the output channel, kernel size, stride, and padding. Compared with the Conv1D module, the ConvTranspose1D contains an extra output padding parameter.}

\begin{adjustbox}{width=.45\textwidth,center}
\begin{tabular}{lcc}
\toprule
Layer name & Structure & Output \\
\hline 
Input & $-$ & $(1, 200)$ \\
\hline
\multicolumn{3}{c}{\textbf{Encoder}} \\
Conv1D + SiLU & $(16, 5, 2, 2)$ & $(16, 100)$\\
Conv1D + SiLU & $(32, 3, 2, 1)$ & $(32, 50)$ \\
Conv1D + SiLU & $(64, 3, 1, 1)$  & $(64, 50)$  \\
Flatten & $-$ & $(3200)$ \\
Linear + LayerNorm & $3200 \rightarrow \textit{latent-dim}$ & $(\textit{latent-dim})$ \\
\hline
\multicolumn{3}{c}{\textbf{Decoder}} \\
LayerNorm + SiLU & $-$ & $(\textit{latent-dim})$ \\
Linear + SiLU & $\textit{latent-dim} \rightarrow 2*\textit{latent-dim}$ & $(2*\textit{latent-dim})$ \\
Linear + SiLU & $2*\textit{latent-dim} \rightarrow 3200$ & $(3200)$ \\
Unflatten & $-$ & $(64, 50)$ \\
ConvTranspose1D + SiLU & $(32, 3, 1, 1, 0)$ & $(32, 50)$ \\
ConvTranspose1D + SiLU & $(16, 3, 2, 1, 1)$ & $(16, 100)$ \\
ConvTranspose1D & $(16, 5, 2, 2, 1)$ & $(1, 200)$ \\
Conv1D & $(16, 5, 1, 2)$ & $(16, 200)$ \\
Conv1D + Sigmoid & $(1, 3, 1, 1)$ & $(1, 200)$ \\
\bottomrule
\end{tabular}
\label{table:efficient_conv_deconv_net}
\end{adjustbox}
\vspace{-10pt}
\end{table}

\subsection{Model Configuration}
\label{ssec:model_configuration}
In our system, we chunk the audio into 16-sec segments as the system input following~\cite{cheng2023target}. Then we extract the 80-dimensional F-bank feature with 10ms frame shift as the acoustic feature.
In our system, we use the ResNet34-based r-vector~\cite{zeinali2019but} pre-trained on Voxceleb2 dataset~\cite{chung2018voxceleb2} to process the acoustic feature first. The frame-level representation before the pooling layer of the r-vector is extracted and fed into the Conformer. Since the r-vector system will downsample the input for 8-times, the length of the output from r-vector is 200. Similarly, we also set the output VAD length $L$ for each segment to 200. For the model configuration of the Conformer and the Decoder, we directly follow the setup in the original Seq2Seq-TSVAD paper~\cite{cheng2023target}. Apart from the TS-VAD system, we also introduced a VAD label auto-encoder (Label-AE), and we show the detailed configuration for the Label-AE in Table \ref{table:efficient_conv_deconv_net}.

\subsection{System Training}
In our experiment, we first pre-train the Label-AE model on the simulation dataset for 20 epochs. Then we train our proposed Flow-TSVAD system. Following the implementation in~\cite{cheng2023target}, we pad the input of the decoder with augmented speaker embeddings to ensure that all inputs have the same number of speakers, which is 20. Each augmented embedding has a 50\% chance of being set to zeros; otherwise, it is replaced by another embedding not present in the current audio. Additionally, there is a 20\% chance that all input speaker embeddings will be replaced with non-existent speakers. We first fix the r-vector and train the system on the simulation dataset for 20 epochs, and then we unfix the r-vector and continue the training for another 30 epochs. Finally, the system with unfixed r-vector is finetuned on the CALLHOME part1 for 50 epochs.

\subsection{Evaluation Setup}
In the evaluation stage, we first trained a clustering-based (Spectral Clustering) diarization system following the wespeaker~\cite{wang2023wespeaker} recipe\footnote{\url{https://github.com/wenet-e2e/wespeaker/tree/master/examples/voxconverse/v1}} to provide the initial diarization results for TS-VAD systems. Based on this initial diarization results, the pre-trained r-vector introduced in section \ref{ssec:model_configuration} is used to extract enrolled speaker embedding. In this process, the speech segments shorter than 2s are abandoned. For the evaluation metric, we calculate the Miss Rate (MS), False Alarm (FA), Confusion Error (Conf.), and Diarization Error Rate (DER) with a collar equal to 250ms.

\section{Results and Analysis}

\subsection{Analysis for the Impact of Different Latent Dimensions}
As introduced in Section \ref{ssec:label_ae}, we hypothesize that applying the FM algorithm in a dense, low-dimensional latent space is more effective than using the original binary label sequence space. In this section, we evaluate the impact of applying FM in the latent space and the influence of different latent dimensions. The results are presented in Table \ref{table:hidden_dim_ablation}.

First, we observe that the Diarization Error Rate (DER) is extremely high when FM is not applied in the latent space, indicating that this approach is ineffective. However, when FM is applied in the latent space, system performance improves significantly. Additionally, we find that the Flow-TSVAD model operating in a 32-dimensional latent space achieves the best results. This is reasonable, as the Label-AE (32) model offers better reconstruction ability than the Label-AE (16) and a more densely packed latent space compared to the Label-AE (64). Consequently, we will use the Label-AE (32) model in subsequent experiments.

\begin{table}[ht!]
\centering
\caption{\textbf{CALLHOME diarization results (\%) with different hidden dimensions.} The Recon. DER (reconstruction DER) means that we directly feed the CALLHOME binary label sequence into our proposed Label-AE in section \ref{ssec:label_ae} and evaluate the reconstruction label. For all the Flow-TSVAD systems, we set the inference step in the FM algorithm to 2.
}
\begin{adjustbox}{width=.43\textwidth,center}
\begin{threeparttable}
\begin{tabular}{l|c|cccc}
\toprule
Latent Dim & Recon. DER  & MS & FA & Conf. & DER \\

\hline
- & - &  1.01 & 78.74 & 2.61 & 82.35 \\
\hline
16 & 1.69 & 6.76  & 2.93 & 2.58 & 12.28 \\
32 & 0.00 & 6.81 & 1.96 & 2.43 & 11.21 \\
64 & 0.00 & 6.92 & 2.30 & 2.35 & 11.56 \\
\bottomrule
\end{tabular}
\end{threeparttable}
\label{table:hidden_dim_ablation}
\end{adjustbox}
\vspace{-5pt}
\end{table}

\subsection{Analysis for Different Inference Steps}
\begin{figure}[ht!]
\vspace{-10pt}
    \centering
    \includegraphics[width=\linewidth]{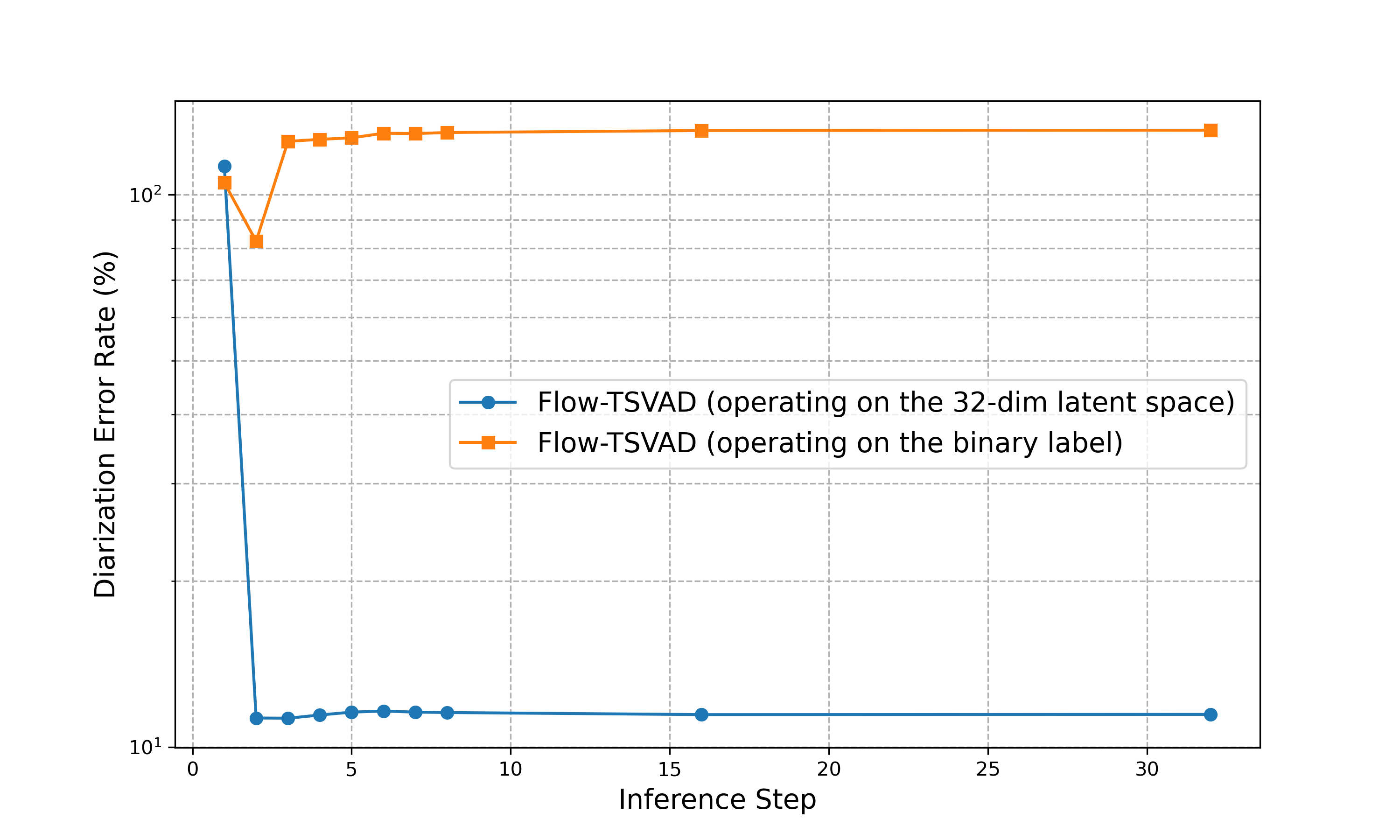}
    \caption{\textbf{The DER (\%) variation for different inference steps.} The results in the figure are infered with the steps 1, 2, 3, 4, 5, 6, 7, 8, 16, 32, respectively.
    }
    \label{fig:der_with_flow_step}
\vspace{-10pt}
\end{figure}

As introduced in Section \ref{ssec:label_ae}, solving the ODE function is necessary to generate the final results in the FM algorithm. To model the ODE path accurately, researchers typically solve this problem iteratively~\cite{lipman2023flow} using specified inference steps. Here, we evaluate our Flow-TSVAD system with different inference steps and present the results in Figure \ref{fig:der_with_flow_step}.

Firstly, we observe that when operating the FM on the binary label sequence, increasing the number of inference steps does not lead to the model's predictions converging to a reasonable value. Interestingly, when applying FM in the latent space, the model's inference converges very quickly, reaching a stable state by the second step, with minimal performance change in subsequent steps. This behavior contrasts with FM in other tasks, such as text-to-speech~\cite{le2024voicebox} and image generation~\cite{lipman2023flow}, which typically require more steps for convergence. This finding enables us to utilize the FM method in speaker diarization tasks efficiently, without the concern of excessively slowing down model inference.

\subsection{Analysis of the Randomness from Different Sampling}
\label{ssec:analysis_different_sampling}

\begin{figure}[ht!]
\vspace{-5pt}
    \centering
    \includegraphics[width=\linewidth]{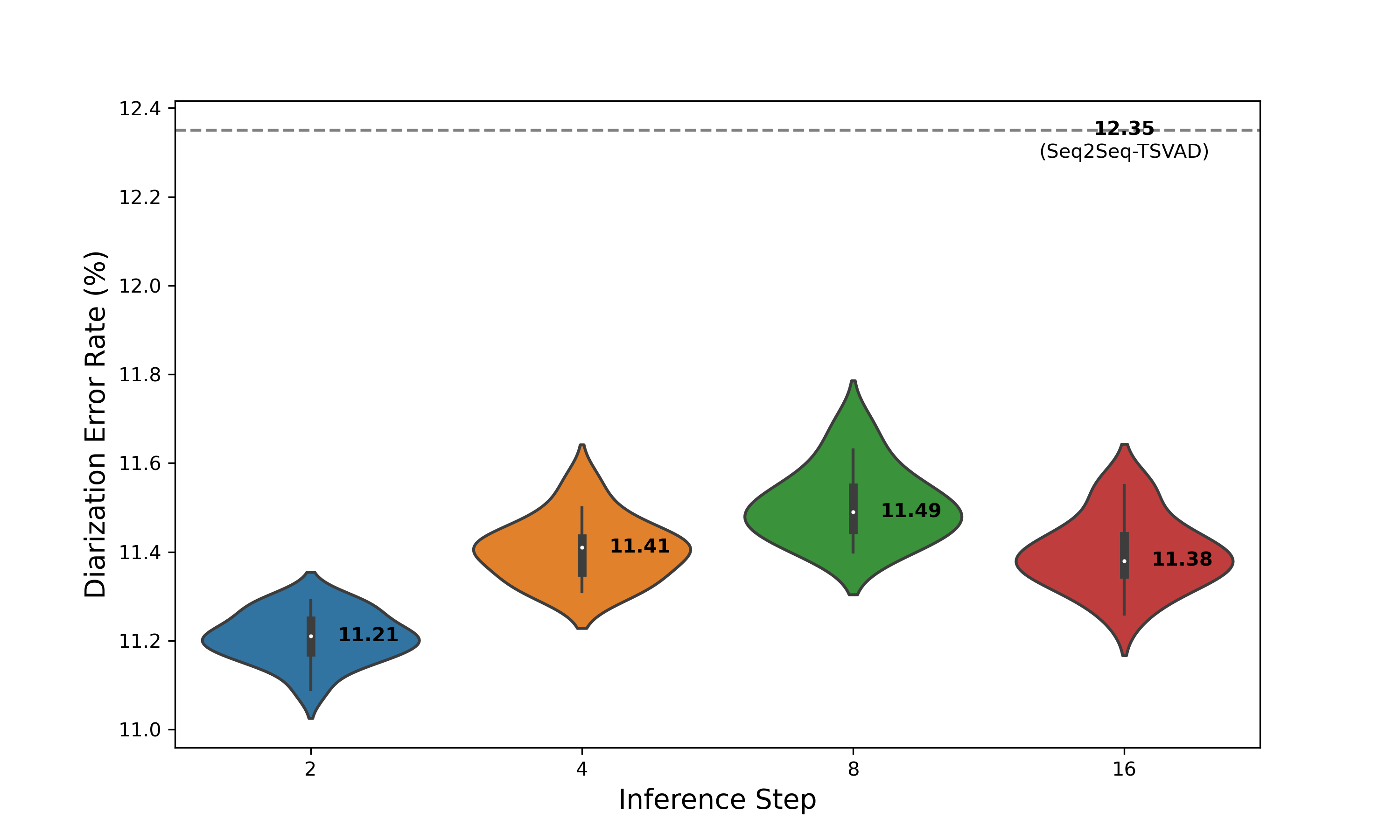}
    \caption{\textbf{The DER (\%) distribution violin plot for different inference steps.} For each inference step, we sample 15 times to generate the diarization results with different random seeds.
    }
    \label{fig:der_distribution}
\vspace{-5pt}
\end{figure}

The primary distinction between generative and discriminative models is that generative models approximate a distribution. During inference, this involves sampling from the distribution, meaning different sampling operations can yield varying results. In this section, we analyze the impact of multiple sampling instances on the results and present the DER (\%) performance distribution in Figure \ref{fig:der_distribution}. The figure demonstrates that running the Flow-TSVAD system multiple times introduces some performance variation. However, this variation is minimal, and all results depicted in the figure outperform the baseline Seq2Seq-TSVAD performance listed in Table \ref{table:res_comparison}.

\subsection{Performance Comparison between Different Systems}
In this section, we compare our proposed Flow-TSVAD system with other diarization systems. The results in Table \ref{table:res_comparison} show that the Flow-TSVAD model operating in the latent space outperforms both the Seq2Seq-TSVAD baseline and other neural network-based systems. Additionally, as discussed in Section \ref{ssec:analysis_different_sampling}, Flow-TSVAD, being a generative method, produces different results with different sampling operations. Notably, if we ensemble the results from various sampling instances using the dover-lap~\cite{Raj2021Doverlap} method, the model's performance improves further.

\begin{table}[ht!]
\vspace{-5pt}
\centering
\caption{\textbf{Results (\%) on CALLHOME part2.} When using the Flow-TSVAD model, we set the inference step to 2. For the Dover-lap Ensemble, we run our systems for 3 times.}
\begin{adjustbox}{width=.43\textwidth,center}
\begin{threeparttable}
\begin{tabular}{l|cccc}
\toprule
System & MS & FA & Conf. & DER \\
\hline
Spectral Cluster & 12.87 & 1.00 & 5.48 & 19.34 \\
Seq2Seq-TSVAD~\cite{cheng2023target}$^\dagger$ & 8.21 & 2.02 & 2.11 & 12.35 \\
USED-F~\cite{ao2023used } & - & - & - & 13.16 \\
AED-EEND~\cite{chen23n_interspeech} & - & - & - & 14.00 \\

EEND-GLA-Large~\cite{horiguchi2022online} & - & - & - & 11.84 \\
\hline
Flow-TSVAD & 6.81 & 1.96 & 2.43 & 11.21 \\
\hspace{3pt} + Dover-lap Ensemble & 6.24 & 2.02 & 2.66 & 10.91 \\

\bottomrule
\end{tabular}
\begin{tablenotes}\footnotesize
\item $^\dagger$: we train the Seq2Seq-TSVAD system following the same training setup as our proposed Flow-TSVAD.
\end{tablenotes}
\end{threeparttable}
\label{table:res_comparison}
\end{adjustbox}
\vspace{-5pt}
\end{table}

\section{Conclusion}
In this paper, we propose leveraging the generative algorithm, Flow-Matching, for the typically discriminative task, speaker diarization, for the first time. Our experiments revealed that applying the FM algorithm directly to the original speaker activity binary label space is ineffective. To address this issue, we introduced a label auto-encoder (Label-AE) to map the binary label sequence into a continuous and denser space. Our proposed Flow-TSVAD system outperforms the baseline Seq2Seq-TSVAD system. Furthermore, we observed that Flow-TSVAD requires only two inference steps to achieve promising results. As a generative model, Flow-TSVAD produces different results when run multiple times, and an ensemble of these different results can yield even better performance. While we have successfully applied a generative algorithm to the TS-VAD paradigm, further research is needed to explore the use of generative algorithms in other diarization paradigms, such as End-to-End Neural Diarization (EEND).

\clearpage
\footnotesize
\bibliographystyle{IEEEbib}
\bibliography{refs}

\begin{thebibliography}{10}

\bibitem{kanda2019simultaneous}
Naoyuki Kanda, Shota Horiguchi, Yusuke Fujita, Yawen Xue, Kenji Nagamatsu, and Shinji Watanabe,
\newblock ``Simultaneous speech recognition and speaker diarization for monaural dialogue recordings with target-speaker acoustic models,''
\newblock in {\em ASRU}. IEEE, 2019, pp. 31--38.

\bibitem{ma2024wenetspeech4tts}
Linhan Ma, Dake Guo, Kun Song, Yuepeng Jiang, Shuai Wang, Liumeng Xue, Weiming Xu, Huan Zhao, Binbin Zhang, and Lei Xie,
\newblock ``Wenetspeech4tts: A 12,800-hour mandarin tts corpus for large speech generation model benchmark,''
\newblock {\em arXiv preprint arXiv:2406.05763}, 2024.

\bibitem{tranter2006overview}
Sue~E Tranter and Douglas~A Reynolds,
\newblock ``An overview of automatic speaker diarization systems,''
\newblock {\em IEEE Transactions on audio, speech, and language processing}, vol. 14, no. 5, pp. 1557--1565, 2006.

\bibitem{wang2018speaker}
Quan Wang, Carlton Downey, Li~Wan, Philip~Andrew Mansfield, and Ignacio~Lopz Moreno,
\newblock ``Speaker diarization with lstm,''
\newblock in {\em ICASSP}. IEEE, 2018, pp. 5239--5243.

\bibitem{valente2010variational}
Fabio Valente, Petr Motlicek, and Deepu Vijayasenan,
\newblock ``Variational bayesian speaker diarization of meeting recordings,''
\newblock in {\em ICASSP}. IEEE, 2010, pp. 4954--4957.

\bibitem{landini2022bayesian}
Federico Landini, J{\'a}n Profant, Mireia Diez, and Luk{\'a}{\v{s}} Burget,
\newblock ``Bayesian hmm clustering of x-vector sequences (vbx) in speaker diarization: theory, implementation and analysis on standard tasks,''
\newblock {\em Computer Speech \& Language}, vol. 71, pp. 101254, 2022.

\bibitem{fujita2019end_lstm}
Yusuke Fujita, Naoyuki Kanda, Shota Horiguchi, Kenji Nagamatsu, and Shinji Watanabe,
\newblock ``{End-to-End Neural Speaker Diarization with Permutation-Free Objectives},''
\newblock in {\em Proc. Interspeech 2019}, 2019, pp. 4300--4304.

\bibitem{fujita2019end_sa}
Yusuke Fujita, Naoyuki Kanda, Shota Horiguchi, Yawen Xue, Kenji Nagamatsu, and Shinji Watanabe,
\newblock ``End-to-end neural speaker diarization with self-attention,''
\newblock in {\em ASRU}. IEEE, 2019, pp. 296--303.

\bibitem{horiguchi2022encoder}
Shota Horiguchi, Yusuke Fujita, Shinji Watanabe, Yawen Xue, and Paola Garcia,
\newblock ``Encoder-decoder based attractors for end-to-end neural diarization,''
\newblock {\em IEEE/ACM Transactions on Audio, Speech, and Language Processing}, vol. 30, pp. 1493--1507, 2022.

\bibitem{chen23n_interspeech}
Zhengyang Chen, Bing Han, Shuai Wang, and Yanmin Qian,
\newblock ``{Attention-based Encoder-Decoder Network for End-to-End Neural Speaker Diarization with Target Speaker Attractor},''
\newblock in {\em Proc. INTERSPEECH 2023}, 2023, pp. 3552--3556.

\bibitem{chen2024attention}
Zhengyang Chen, Bing Han, Shuai Wang, and Yanmin Qian,
\newblock ``Attention-based encoder-decoder end-to-end neural diarization with embedding enhancer,''
\newblock {\em IEEE/ACM Transactions on Audio, Speech, and Language Processing}, vol. 32, pp. 1636--1649, 2024.

\bibitem{medennikov2020target}
Ivan Medennikov, Maxim Korenevsky, Tatiana Prisyach, Yuri Khokhlov, Mariya Korenevskaya, Ivan Sorokin, Tatiana Timofeeva, Anton Mitrofanov, Andrei Andrusenko, Ivan Podluzhny, Aleksandr Laptev, and Aleksei Romanenko,
\newblock ``{Target-Speaker Voice Activity Detection: A Novel Approach for Multi-Speaker Diarization in a Dinner Party Scenario},''
\newblock in {\em Proc. Interspeech 2020}, 2020, pp. 274--278.

\bibitem{cheng2023target}
Ming Cheng, Weiqing Wang, Yucong Zhang, Xiaoyi Qin, and Ming Li,
\newblock ``Target-speaker voice activity detection via sequence-to-sequence prediction,''
\newblock in {\em ICASSP}. IEEE, 2023, pp. 1--5.

\bibitem{cheng2024multi}
Ming Cheng and Ming Li,
\newblock ``Multi-input multi-output target-speaker voice activity detection for unified, flexible, and robust audio-visual speaker diarization,''
\newblock {\em arXiv preprint arXiv:2401.08052}, 2024.

\bibitem{medennikov2020stc}
Ivan Medennikov, Maxim Korenevsky, Tatiana Prisyach, Yuri Khokhlov, Mariya Korenevskaya, Ivan Sorokin, Tatiana Timofeeva, Anton Mitrofanov, Andrei Andrusenko, Ivan Podluzhny, et~al.,
\newblock ``The stc system for the chime-6 challenge,''
\newblock in {\em CHiME 2020 Workshop on Speech Processing in Everyday Environments}, 2020.

\bibitem{wang2022dku}
Weiqing Wang, Xiaoyi Qin, Ming Cheng, Yucong Zhang, Kangyue Wang, and Ming Li,
\newblock ``The dku-smiip diarization system for the voxceleb speaker recognition challenge 2022,''
\newblock in {\em Voxsrc Workshop}, 2022.

\bibitem{richter2023speech}
Julius Richter, Simon Welker, Jean-Marie Lemercier, Bunlong Lay, and Timo Gerkmann,
\newblock ``Speech enhancement and dereverberation with diffusion-based generative models,''
\newblock {\em IEEE/ACM Transactions on Audio, Speech, and Language Processing}, vol. 31, pp. 2351--2364, 2023.

\bibitem{welker22_interspeech}
Simon Welker, Julius Richter, and Timo Gerkmann,
\newblock ``{Speech Enhancement with Score-Based Generative Models in the Complex STFT Domain},''
\newblock in {\em Proc. Interspeech 2022}, 2022, pp. 2928--2932.

\bibitem{kamo2023target}
Naoyuki Kamo, Marc Delcroix, and Tomohiro Nakatan,
\newblock ``Target speech extraction with conditional diffusion model,''
\newblock {\em arXiv preprint arXiv:2308.03987}, 2023.

\bibitem{chen2023diffusiondet}
Shoufa Chen, Peize Sun, Yibing Song, and Ping Luo,
\newblock ``Diffusiondet: Diffusion model for object detection,''
\newblock in {\em CVPR}, 2023, pp. 19830--19843.

\bibitem{gu2024diffusioninst}
Zhangxuan Gu, Haoxing Chen, and Zhuoer Xu,
\newblock ``Diffusioninst: Diffusion model for instance segmentation,''
\newblock in {\em ICASSP}. IEEE, 2024, pp. 2730--2734.

\bibitem{lipman2023flow}
Yaron Lipman, Ricky T.~Q. Chen, Heli Ben{-}Hamu, Maximilian Nickel, and Matthew Le,
\newblock ``Flow matching for generative modeling,''
\newblock in {\em The Eleventh International Conference on Learning Representations, {ICLR} 2023, Kigali, Rwanda, May 1-5, 2023}, 2023.

\bibitem{zeinali2019but}
Hossein Zeinali, Shuai Wang, Anna Silnova, Pavel Mat{\v{e}}jka, and Old{\v{r}}ich Plchot,
\newblock ``But system description to voxceleb speaker recognition challenge 2019,''
\newblock {\em arXiv preprint arXiv:1910.12592}, 2019.

\bibitem{vaswani2017attention}
Ashish Vaswani, Noam Shazeer, Niki Parmar, Jakob Uszkoreit, Llion Jones, Aidan~N Gomez, {\L}ukasz Kaiser, and Illia Polosukhin,
\newblock ``Attention is all you need,''
\newblock {\em NeurlPS}, vol. 30, 2017.

\bibitem{gulati2020conformer}
Anmol Gulati, James Qin, Chung-Cheng Chiu, Niki Parmar, Yu~Zhang, Jiahui Yu, Wei Han, Shibo Wang, Zhengdong Zhang, Yonghui Wu, et~al.,
\newblock ``Conformer: Convolution-augmented transformer for speech recognition,''
\newblock {\em arXiv preprint arXiv:2005.08100}, 2020.

\bibitem{huang2017arbitrary}
Xun Huang and Serge Belongie,
\newblock ``Arbitrary style transfer in real-time with adaptive instance normalization,''
\newblock in {\em ICCV}, 2017, pp. 1501--1510.

\bibitem{callhome}
``2000 nist speaker recognition evaluation,''
\newblock 2000,
\newblock Last accessed on 2023-08-06.

\bibitem{landini22_interspeech}
Federico Landini, Alicia Lozano-Diez, Mireia Diez, and Lukáš Burget,
\newblock ``{From Simulated Mixtures to Simulated Conversations as Training Data for End-to-End Neural Diarization},''
\newblock in {\em Proc. Interspeech 2022}, 2022, pp. 5095--5099.

\bibitem{chung2018voxceleb2}
Joon~Son Chung, Arsha Nagrani, and Andrew Zisserman,
\newblock ``Voxceleb2: Deep speaker recognition,''
\newblock {\em arXiv preprint arXiv:1806.05622}, 2018.

\bibitem{wang2023wespeaker}
Hongji Wang, Chengdong Liang, Shuai Wang, Zhengyang Chen, Binbin Zhang, Xu~Xiang, Yanlei Deng, and Yanmin Qian,
\newblock ``Wespeaker: A research and production oriented speaker embedding learning toolkit,''
\newblock in {\em ICASSP}. IEEE, 2023, pp. 1--5.

\bibitem{le2024voicebox}
Matthew Le, Apoorv Vyas, Bowen Shi, Brian Karrer, Leda Sari, Rashel Moritz, Mary Williamson, Vimal Manohar, Yossi Adi, Jay Mahadeokar, et~al.,
\newblock ``Voicebox: Text-guided multilingual universal speech generation at scale,''
\newblock {\em NeurlPS}, vol. 36, 2024.

\bibitem{Raj2021Doverlap}
D.Raj, P.Garcia, Z.Huang, S.Watanabe, D.Povey, A.Stolcke, and S.Khudanpur,
\newblock ``{DOVER-Lap}: A method for combining overlap-aware diarization outputs,''
\newblock {\em 2021 IEEE Spoken Language Technology Workshop (SLT)}, 2021.

\bibitem{ao2023used}
Junyi Ao, Mehmet~Sinan Y{\i}ld{\i}r{\i}m, Meng Ge, Shuai Wang, Ruijie Tao, Yanmin Qian, Liqun Deng, Longshuai Xiao, and Haizhou Li,
\newblock ``Used: Universal speaker extraction and diarization,''
\newblock {\em arXiv preprint arXiv:2309.10674}, 2023.

\bibitem{horiguchi2022online}
Shota Horiguchi, Shinji Watanabe, Paola Garc{\'\i}a, Yuki Takashima, and Yohei Kawaguchi,
\newblock ``Online neural diarization of unlimited numbers of speakers using global and local attractors,''
\newblock {\em IEEE/ACM Transactions on Audio, Speech, and Language Processing}, 2022.

\end{thebibliography}
\end{document}